\documentclass[12pt]{iopart}
\usepackage{iopams}
\begin{document}

\title[Exact partition functions of the Ising model on $M\times N$ planar lattices]
{Exact partition functions of the Ising model on $M\times N$
planar lattices with periodic-aperiodic boundary conditions}

\author{Ming-Chya Wu\dag\footnote[3]{Institute of Physics, Academia Sinica, Nankang,
Taipei 11529, Taiwan, Republic of China.\\
E-mail: mcwu@phys.sinica.edu.tw\\
Tel: 886-2-2789-6762\\
Fax: 886-2-2782-2467}\ and Chin-Kun Hu\dag\footnote[7]{E-mail:
huck@phys.sinica.edu.tw}}

\address{\dag Institute of Physics, Academia Sinica, Nankang, Taipei 11529,
Taiwan, Republic of China}


\begin{abstract}
The Grassmann path integral approach is used to calculate exact
partition functions of the Ising model on $M\times N$ square (sq),
plane triangular (pt) and honeycomb (hc) lattices with
periodic-periodic (pp), periodic-antiperiodic (pa),
antiperiodic-periodic (ap) and antiperiodic-antiperiodic (aa)
boundary conditions. The partition functions are used to calculate
and plot the specific heat, $C/k_B$, as a function of the
temperature, $\theta =k_BT/J$. We find that for the $N\times N$ sq
lattice, $C/k_B$ for pa and ap boundary conditions are different
from those for aa boundary conditions, but for the $N\times N$ pt
and hc lattices, $C/k_B$ for ap, pa, and aa boundary conditions
have the same values. Our exact partition functions might also be
useful for understanding the effects of lattice structures and
boundary conditions on critical finite-size corrections of the
Ising model.
\end{abstract}


\maketitle

\section{Introduction}

Universality and scaling are two important concepts in the theory of
critical phenomena \cite{stanley71,kadanoff90} and the Ising model \cite
{onsager} has been a model widely used in such studies. Recently, exact
universal amplitude ratios and finite-size corrections to scaling in
critical Ising model on planar lattices have received much attention \cite
{mcwu,ckhupre99,queiroz,ivashkevich,izmailian,izmailianletter,salas,jankeprb,izmailianep}%
. This may be due to the fact that the hypothesis of universality
leads naturally to the consideration of universal critical
amplitudes and amplitude combinations \cite{barber}, and for the
comparison between experiment and theory in relation to scaling
and universality, it is often a more rigorous test to use
amplitude relations rather than critical exponent values.
Moreover, it is also well known that the finite-size scaling
functions depend on the boundary conditions \cite{huck94}, and
there has been considerable recent interest in studying lattice
model with various boundary conditions
\cite{hu,ckhuprl96,okabe,fywu,ko,okabepre99,tomita}. The study of
exact universal amplitude ratios and finite-size corrections to
scaling in critical Ising model is usually based on the analytical
solutions of the model on finite lattices. Although the exact
solution of the Ising model on $M\times N$ square (sq) lattice had
been obtained long time ago \cite{kaufman}, and the exact
expression of the partition function of the Ising model on
$M\times N$ plane triangular (pt) lattice has been obtained by
lattice field theories recently \cite{nash}, there is still no
published results for the exact solutions of the Ising model on
$M\times N$ pt and honeycomb (hc) lattices with periodic-aperiodic
boundary conditions. The purpose of this paper is to fill this
gap. In the present paper we use the Grassmann path integral to
calculate exact partition functions of the Ising model on $M\times
N$ sq, pt and hc lattices with periodic-periodic (pp),
periodic-antiperiodic (pa), antiperiodic-periodic (ap) and
antiperiodic-antiperiodic (aa) boundary conditions. The partition
functions are used to calculate and plot the specific heat,
$C/k_B$, as a function of the temperature, $\theta =k_BT/J$. We
find that for the $N\times N$ sq lattice, $C/k_B$ for pa and ap
boundary conditions are different from those
for aa boundary conditions, but for the $N\times N$ pt and hc lattices, $%
C/k_B$ for ap, pa, and aa boundary conditions have the same values. Our
exact partition functions might also be useful for understanding the effects
of lattice structures and boundary conditions on critical finite-size
corrections of the Ising model.

Two-dimensional Ising model on the sq lattice at vanishing magnetic field
was first solved by Onsager by the use of Lie algebra \cite{onsager}. The
exact solution he obtained was Ising model on an infinite lattice. The
original method was rather complicated, and it was later improved by Kaufman
\cite{kaufman} who obtained the exact solution of the Ising model on a
finite torus by using the theory of spinor representation. The successful
treatments of the two-dimensional Ising model brought the studies of phase
transition into the modern era. Onsager's solution in one hand showed the
previous classical theories were unreliable in their quantitative
predictions, and on the other hand provided a great stimulus to explore the
true behaviour near the critical point. After Onsager's original solution,
many quite different mathematical approaches were developed, but the
approaches were still complicated. Among them, Schultz, Mattis and Lieb gave
explicitly the fermionic treatment in the framework of transfer-matrix
formalism \cite{shultz}, and Kac and Ward developed the combinatorial method
\cite{kac,green}. Both methods reformulated the two-dimensional Ising model
as a free-fermionic field theory in terms of anticommuting Grassmann
variables, which enclosed the fact that the Ising model on two dimensional
regular lattices may be viewed as free-fermionic theory. The other
alternative method in literature was the Pfaffian representation, which was
introduced by Kasteleyn \cite{kasteleyn} to translate Ising spins into
dimers that can be reduced to some Pfaffian \cite{fisherjmp}. Stephenson has
used the Pfaffian representation to solve the Ising model on the pt lattice,
but the solution was restricted to $6L\times 6L$ lattice due to its $6\times
6$ basic nonvanishing matrix elements and was exact only in the limit of $%
L\rightarrow \infty $ \cite{stephenson}. Recently, by using the connections
between Pfaffian, dimer and Ising model, Nash and O'Connor have obtained the
exact expression of the partition function of the pt lattice Ising model on
a finite torus \cite{nash}. They first employed the lattice field theories
to obtain the exact partition function of the Gaussian model, and then
established the exact expression of the partition function of the pt lattice
Ising model from the analysis of the appropriate lattice determinants and
the parameterization according to the results in \cite{stephenson}.

On the other hand, in view of the simplifying the approach, a remarkable
progress was achieved by Plechko who modified the traditional fermionic
interpretation and introduced a nonstandard approach \cite{plechko1}. By the
use of this approach, Plechko himself has not only rederived Onsager's and
Kaufman's results in a relatively simple way \cite{plechko1}, but also
obtained the partition functions of a class of triangular type decorated
lattices \cite{plechko2}, and a triangular lattice net with holes \cite
{plechko3}. Quite recently, by using the same approach, Wu \textit{at al}.
have obtained the $M\times N$ sq lattice Ising model with periodic-aperiodic
boundary condition \cite{mcwu}, and Liaw \textit{at al}. have successfully
solved triangular and hexagonal lattices on a cylinder geometry ($M\times
\infty $) with periodic and antiperiodic boundary condition \cite{liaw}.
This approach is based on the integration over the anticommuting Grassmann
variables and the mirror-ordered factorization principle in two-dimensional
density matrix \cite{plechko1,plechko2,plechko3,liaw}, and does not involve
the traditional transfer-matrix or combinatorial considerations. The whole
scheme of the method can be illustrated schematically as shown below \cite
{plechko1}:
\[
Z=\mathop{\rm Sp}\limits_{\left( \sigma \right) }\left\{ Z\left( \sigma
\right) \right\} \rightarrow \mathop{\rm Sp}\limits_{\left( \sigma \left|
\chi \right. \right) }\left\{ Z\left( \sigma \left| \chi \right. \right)
\right\} \rightarrow \mathop{\rm Sp}\limits_{\left( \chi \right) }\left\{
Z\left( \chi \right) \right\} =Z,
\]
where $^{\backprime \backprime }\mathrm{Sp}^{\prime \prime }$ stands for the
average over spin variables ($\sigma $) or Grassmann variables ($\chi $).
The original partition function $Z$ is expressed purely by spin variables ($%
\sigma $) at each lattice site. With a set of anticommuting Grassmann
variables ($\chi $) being introduced to factorize the local bond Boltzmann
weight such that spin variables are decoupled, the partition function passes
to a mixed $Z\left( \sigma \left| \chi \right. \right) $ representation.
Then, by eliminating the spin variables in the mixed $Z\left( \sigma \left|
\chi \right. \right) $ representation, the fermionic interpretation $Z\left(
\chi \right) $ of the two-dimensional Ising model can be obtained, and after
carrying out the Grassmann integral, the analytic solution for the partition
function and free energy can be achieved \cite
{plechko1,plechko2,plechko3,liaw}.

In the present paper, we work in this framework to obtain exact partition
functions of $M\times N$ pt and hc lattices with different boundary
conditions, including pp, pa, ap and aa boundary conditions. We used these
results to calculate and plot the specific heat, $C/k_B$, as a function of
the temperature, $\theta =k_BT/J$. Our results show that for the sq lattice,
$C/k_B$ for pa and ap boundary conditions are different from those for aa
boundary conditions, but for the pt and hc lattices, $C/k_B$ for ap, pa, and
aa boundary conditions have the same values. Beside these analyses, our
exact partition functions may also be used for understanding the effects of
lattice structures and boundary conditions on critical properties and
critical finite-size corrections of the Ising model.

This paper is organized as follows. In section 2, we set up a general form
of the partition function for pt and hc lattices. Then, three pairs of
conjugate Grassmann variables are introduced for a lattice site to factorize
the Boltzmann weights, and the principle of mirror ordering are used to
rearrange the Grassmann factors so we can perform the summation over Ising
spins to obtain a pure fermionic expression of the partition function. In
section 3, using the Fourier transform technique we complete the
integrations over the Grassmann variables to obtain the exact solution of
the partition function. Then, the solution is subjected to
periodic-aperiodic boundary conditions, including pp, pa, ap and aa boundary
conditions. We further consider the shift behaviours of the maximum of the
specific heats of these systems in section 4. Finally, we discuss some
problems for further studies in section 5.

\section{The Partition Function}

Consider Ising ferromagnets on $M\times N$ pt and hc lattices as shown in
figure 1, in which the former is considered as a sq lattice with a single
second-neighbor interaction, and the latter contains an inner spin in each
lattice cell. The corresponding Hamiltonians, respectively, read as
\begin{equation}
H_t=-\sum_{m=1}^M\sum_{n=1}^N\left( J_1\sigma _{mn}\sigma _{m+1n}+J_2\sigma
_{mn}\sigma _{mn+1}+J_3\sigma _{m+1n}\sigma _{mn+1}\right) ,
\end{equation}
and
\begin{equation}
H_h=-\sum_{m=1}^M\sum_{n=1}^N\left( J_1\sigma _0\sigma _{mn}+J_2\sigma
_0\sigma _{mn+1}+J_3\sigma _0\sigma _{m+1n}\right) ,
\end{equation}
where $J_i$ with $i=1,2,3$ are the coupling constants ($J_i>0$ for
ferromagnetic lattices), $\sigma _{mn}=\pm 1$ is the Ising spin located at
the site $(m,n)$, and $\sigma _0$ denotes the inner Ising spin in hc
lattice. Using the identity of the Boltzmann weight,
\begin{equation}
\exp \left( \beta J_i\sigma _\mu \sigma _\nu \right) =\cosh \left( \beta
J_i\right) \left[ 1+\tanh \left( \beta J_i\right) \sigma _\mu \sigma _\nu
\right] ,
\end{equation}
$\beta =\left( k_BT\right) ^{-1}$, and performing the sum over $\sigma _0$,
the partition functions of two lattices can be formulated in a single three
spin-polynomial representation,
\begin{eqnarray}
\fl Z &=&2^{N_s}\left[ \prod_{i=1}^{n_b}\cosh \left( \beta J_i\right)
\right] ^{N_s}  \nonumber \\
\fl && \times \mathop{\rm Sp}\limits_{\left( \sigma \right) }\left\{
\prod_{m=1}^M\prod_{n=1}^N\left( \alpha _0+\alpha _1\sigma _{mn}\sigma
_{m+1n}+\alpha _2\sigma _{mn}\sigma _{mn+1}+\alpha _3\sigma _{m+1n}\sigma
_{mn+1}\right) \right\} ,
\end{eqnarray}
where $N_s$ is the numbers of lattice sites ($N_s=MN$ for sq and pt lattice,
$N_s=2MN$ for hc lattice) and $n_b$ is the number of bonds per lattice cell (%
$n_b=2$ for sq lattice, $n_b=3$ for pt and hc lattice), symbol $^{\backprime
\backprime }\mathop{\rm Sp}\limits_{\left( \sigma \right) }$ $^{\prime
\prime }$ stands for spin average defined by
\begin{equation}
\mathop{\rm Sp}\limits_{\left( \sigma _i\right) }\left[ \cdots \right]
=\frac 12\sum_{\left( \sigma _i=\pm 1\right) }\left[ \cdots \right] ,\quad %
\mathop{\rm Sp}\limits_{\left( \sigma _i\right) }\left[ 1\right] =1,\quad %
\mathop{\rm Sp}\limits_{\left( \sigma _i\right) }\left[ \sigma _i\right] =0
\end{equation}
and $\alpha _i$'s are defined as
\begin{equation}
\alpha _0^T=1+t_1t_2t_3,\ \alpha _1^T=t_1+t_2t_3,\ \alpha _2^T=t_2+t_3t_1,\
\alpha _3^T=t_3+t_1t_2,  \label{tricoef}
\end{equation}
$t_i=\tanh \left( \beta J_i\right) $ with $i=1,2,3$, for pt lattice, and
\begin{equation}
\alpha _0^H=1,\ \alpha _1^H=t_1t_3,\ \alpha _2^H=t_1t_2,\ \alpha _3^H=t_2t_3,
\label{honcoef}
\end{equation}
for hc lattice.

To factorize the partition, we rewrite the partition function as
\begin{eqnarray}
\fl Z_H &=&2^{N_s}\left[ \prod_{i=1}^{n_b}\cosh \left( \beta J_i\right)
\right] ^{N_s}  \nonumber \\
\fl &&\times \mathop{\rm Sp}\limits_{\left( \sigma \right) }\left\{
\prod_{m=1}^M\prod_{n=1}^Nr_0\left( 1+r_1\sigma _{mn}\sigma _{m+1n}\right)
\left( 1+r_2\sigma _{mn}\sigma _{mn+1}\right) \left( 1+r_3\sigma
_{m+1n}\sigma _{mn+1}\right) \right\} ,
\end{eqnarray}
where $r_i$ with $i=0,1,2,3$ vary from one lattice to the other, and are
related to $\alpha _i$'s from
\begin{equation}
\fl \alpha _0=r_0\left( 1+r_1r_2r_3\right) ,\alpha _1=r_0\left(
r_1+r_2r_3\right) ,\alpha _2=r_0\left( r_2+r_1r_3\right) ,\alpha
_3=r_0\left( r_3+r_1r_2\right) .  \label{coef}
\end{equation}
For pt lattice, the relation between $r_i$ and $t_i$ is trivial, i.e. $r_0=1$
and $r_i=t_i$, but for hc lattice, the relation is nontrivial and is
determined by equations (\ref{honcoef}) and (\ref{coef}).

It is more convenient to define the generalized reduced partition function
as
\begin{equation}
Q=r_0^{MN}\tilde Q,
\end{equation}
with
\begin{equation}
\fl \tilde Q=\prod_{m=1}^M\prod_{n=1}^N\mathop{\rm Sp}\limits_{\left( \sigma
_{mn}\right) }\left[ \left( 1+r_1\sigma _{mn}\sigma _{m+1n}\right) \left(
1+r_2\sigma _{mn}\sigma _{mn+1}\right) \left( 1+r_3\sigma _{mn+1}\sigma
_{m+1n}\right) \right] .
\end{equation}

To construct the fermionic representation of the generalized partition
function, we associate each lattice site $\left( m,n\right) $ with three
pairs of conjugate Grassmann variables, $\left\{
a_{mn},a_{mn}^{*};b_{mn},b_{mn}^{*};c_{mn},c_{mn}^{*}\right\} \in \chi $.
All of these Grassmann variables are anticommuting, and their square are
zero. Their integral obeys the basic rules \cite{berezin}
\begin{equation}
\int d\chi =0,\quad \int d\chi \cdot \chi =1,
\end{equation}
\begin{equation}
\int d\chi \cdot \Omega \left( \chi +\eta \right) =\int d\chi \cdot \Omega
\left( \chi \right) ,
\end{equation}
for an arbitrary vector $\eta $ with anticommuting components, and there is
the relation
\begin{equation}
1+r_i\sigma _\mu \sigma _\nu =\int d\chi ^{*}d\chi e^{\chi \chi ^{*}}\left(
1+\chi \sigma _\mu \right) \left( 1+r_i\chi ^{*}\sigma _\nu \right) .
\end{equation}
Using these Grassmann variables, we can rewrite the reduced partition
function as\cite{plechko1}
\begin{equation}
\fl \tilde Q=\prod_{m=1}^M\prod_{n=1}^N\mathop{\rm Sp}\limits_{\left( \sigma
_{mn}\right) }\left[ \mathop{\rm Sp}\limits_{\left(
a_{mn},b_{mn},c_{mn}\right) }\left(
A_{mn}A_{m+1n}^{*}B_{mn}B_{mn+1}^{*}C_{mn+1}C_{m+1n}^{*}\right) \right] ,
\end{equation}
where $^{\backprime \backprime }\mathop{\rm Sp}\limits_{\left( \chi
_i\right) }$ $^{\prime \prime }$ stands for the averaging with Gaussian
weight
\begin{eqnarray}
\mathop{\rm Sp}\limits_{\left( \chi _i\right) }\left[ \cdots \right] =\int
d\chi _i^{*}d\chi _ie^{\chi _i\chi _i^{*}}\left[ \cdots \right] ,
\end{eqnarray}
with the rules
\begin{eqnarray}
\mathop{\rm Sp}\limits_{\left( \chi _i\right) }\left[ \chi _i\chi
_i^{*}\right] &=&-\mathop{\rm Sp}\limits_{\left( \chi _i\right) }\left[ \chi
_i^{*}\chi _i\right] =1, \\
\mathop{\rm Sp}\limits_{\left( \chi _i\right) }\left[ \chi _i\right] &=&%
\mathop{\rm Sp}\limits_{\left( \chi _i\right) }\left[ \chi _i^{*}\right] =0,
\end{eqnarray}
and the Grassmann factors, $A,A^{*},B,B^{*},C,$ and $C^{*}$, are defined as
\begin{equation}
A_{mn}=1+a_{mn}\sigma _{mn},\quad A_{mn}^{*}=1+r_1a_{m-1n}^{*}\sigma _{mn};
\end{equation}
\begin{equation}
B_{mn}=1+b_{mn}\sigma _{mn},\quad B_{mn}^{*}=1+r_2b_{mn-1}^{*}\sigma _{mn};
\end{equation}
\begin{equation}
C_{mn}=1+c_{mn-1}\sigma _{mn},\quad C_{mn}^{*}=1+r_3c_{m-1n}^{*}\sigma _{mn}.
\end{equation}
In this way, a Boltzmann weight is decoupled to the product of two factors
of separated spins.

For simplicity, we express the reduced partition function as
\begin{equation}
\tilde Q=\mathop{\rm Sp}\limits_{\left( a,b,c\right) }\left\{
\prod_{m=1}^M\prod_{n=1}^N\Psi _{mn}^A\Psi _{mn}^B\Psi _{mn}^C\right\} ,
\end{equation}
where $\Psi _{mn}^A$, $\Psi _{mn}^B$ and $\Psi _{mn}^C$ are defined by
\begin{eqnarray}
\Psi _{mn}^A &=&\mathop{\rm Sp}\limits_{\left( \sigma _{mn}\right) }\left(
A_{mn}A_{m+1n}^{*}\right) , \\
\Psi _{mn}^B &=&\mathop{\rm Sp}\limits_{\left( \sigma _{mn}\right) }\left(
B_{mn}B_{mn+1}^{*}\right) , \\
\Psi _{mn}^C &=&\mathop{\rm Sp}\limits_{\left( \sigma _{mn}\right) }\left(
C_{mn+1}C_{m+1n}^{*}\right) .
\end{eqnarray}

We first treat the boundary weight and consider periodic boundary condition
in both directions:
\begin{eqnarray}
\Psi _{Mn}^A &=&\mathop{\rm Sp}\limits_{\left( \sigma _{Mn}\right) }\left[
\left( 1+a_{Mn}\sigma _{Mn}\right) \left( 1+r_1a_{Mn}^{*}\sigma
_{M+1n}\right) \right]  \nonumber \\
&=&\mathop{\rm Sp}\limits_{\left( \sigma _{Mn}\right) }\left[ \left(
1+r_1a_{0n}^{*}\sigma _{1n}\right) \left( 1+a_{Mn}\sigma _{Mn}\right) \right]
\nonumber \\
&=&\mathop{\rm Sp}\limits_{\left( \sigma _{Mn}\right) }\left(
A_{1n}^{*}A_{Mn}\right) ,
\end{eqnarray}
which implies
\begin{equation}
a_{0n}^{*}=-a_{Mn}^{*}.
\end{equation}
Similarly, from
\begin{eqnarray}
\Psi _{mN}^B &=&\mathop{\rm Sp}\limits_{\left( \sigma _{mN}\right) }\left(
B_{mN}B_{mN+1}^{*}\right) =\mathop{\rm Sp}\limits_{\left( \sigma
_{Nn}\right) }\left( B_{m1}^{*}B_{mN}\right) ,  \label{b1} \\
\Psi _{Mn}^C &=&\mathop{\rm Sp}\limits_{\left( \sigma _{Mn}\right) }\left(
C_{Mn+1}C_{M+1n}^{*}\right) =\mathop{\rm Sp}\limits_{\left( \sigma
_{Mn}\right) }\left( C_{1n}^{*}C_{Mn+1}\right) ,  \label{b2} \\
\Psi _{mN}^C &=&\mathop{\rm Sp}\limits_{\left( \sigma _{mN}\right) }\left(
C_{mN+1}C_{m+1N}^{*}\right) =\mathop{\rm Sp}\limits_{\left( \sigma
_{mN}\right) }\left( C_{m1}C_{m+1N}^{*}\right) ,  \label{b3}
\end{eqnarray}
we have
\begin{eqnarray}
b_{m0}^{*} &=&-b_{mN}^{*}, \\
c_{0n}^{*} &=&-c_{Mn}^{*}, \\
c_{m0} &=&c_{mN}.  \label{am1}
\end{eqnarray}
Since $c_{m0}=c_{mN}$, $\Psi _{mN}^C$ need not to be treated as a boundary
weight, and only $\Psi _{Mn}^A$, $\Psi _{mN}^B$ and $\Psi _{Mn}^C$ should be
considered. However, this situation becomes ambiguous when we take Fourier
transform of these Grassmann variables with single set of exponential
factors in equations (\ref{fourier1}) and (\ref{fourier2}). Because the
Fourier exponential factors are associated with directions in $M$ and $N$,
the sign factor in front of $b_{mN}^{*}$ takes effects simultaneously on $%
b_{mN}^{*}$ and $c_{mN}$. Therefore, the real situation is that instead of
the relation in equation (\ref{am1}), we must take
\begin{equation}
c_{m0}=-c_{mN}.  \label{am2}
\end{equation}
A self-consistent way to assign a minus sign to $c_{m0}$ and obtain the
relation in equation (\ref{am2}) is interchanging $C_{m1}$ in equation (\ref
{b3}) with another Grassmann factor. An equivalent but more convenient
approach is to consider the rearrangement of $B_{m1}^{*}$ in the boundary
weight together with the rearrangement of $C_{m1}$ in the reduced partition
function. To see this, we express the reduced partition function as
\begin{equation}
\fl \tilde Q=\mathop{\rm Sp}\limits_{\left( a,b,c\right) }\left\{ %
\mathop{\rm Sp}\limits_{\left( \sigma \right) }\left[ \left(
\prod_{m=1}^{M-1}\prod_{n=1}^N\Psi _{mn}^A\Psi _{mn}^C\right) \cdot \Psi
_B\cdot \left( \prod_{m=1}^M\prod_{n=1}^{N-1}B_{mn}B_{mn+1}^{*}\right)
\right] \right\} .
\end{equation}
with the boundary weight $\Psi _B$
\begin{eqnarray}
\fl \Psi _B &=&\mathop{\rm Sp}\limits_{\left( a,b,c\right) }\left[ \left(
\prod_{n=1}^N\Psi _{Mn}^A\right) \left( \prod_{m=1}^M\Psi _{mN}^B\right)
\left( \prod_{n=1}^N\Psi _{Mn}^C\right) \right]  \nonumber \\
\fl &=&\mathop{\rm Sp}\limits_{\left( a,b,c\right) }\left[ \left(
\prod_{m=1}^M\stackrel{m}{\overrightarrow{B_{m1}^{*}}}\right) \left(
\prod_{n=1}^N\stackrel{n}{\overleftarrow{C_{1n}^{*}A_{1n}^{*}}}\right)
A_{M1}\left( \prod_{n=2}^N\stackrel{n}{\overrightarrow{C_{Mn}A_{Mn}}}\right)
C_{M1}\prod_{m=1}^M\stackrel{m}{\overleftarrow{B_{mN}}}\right] ,  \label{bw}
\end{eqnarray}
and
\begin{eqnarray}
\fl \prod_{m=1}^{M-1}\prod_{n=1}^N\Psi _{mn}^A\Psi _{mn}^C
&=&\prod_{m=1}^{M-1}\prod_{n=1}^NA_{mn}C_{mn+1}C_{m+1n}^{*}A_{m+1n}^{*}
\nonumber \\
\fl &=&\prod_{m=1}^{M-1}A_{m1}\left( \prod_{n=2}^N\stackrel{n}{%
\overleftarrow{C_{mn}A_{mn}}}\right) C_{m1}\left( \prod_{n=1}^N\stackrel{n}{%
\overrightarrow{C_{m+1n}^{*}A_{m+1n}^{*}}}\right) .  \label{gw}
\end{eqnarray}
Here, arrows have been used to indicate the orders of the products in $m$
and $n$. When we move $B_{m1}^{*}$ from left of $\prod\limits_{n=1}^N%
\stackrel{n}{\overleftarrow{C_{1n}^{*}A_{1n}^{*}}}$ to right of $%
\prod\limits_{n=1}^N\stackrel{n}{\overleftarrow{C_{1n}^{*}A_{1n}^{*}}}$ in
equation (\ref{bw}), $B_{m1}^{*}$ passes $2N$ Grassmann factors, but for
moving $C_{m1}$ from right of $\prod\limits_{n=2}^N\stackrel{n}{%
\overleftarrow{C_{mn}A_{mn}}}$ to left of $A_{m1}$ in equations (\ref{bw})
and (\ref{gw}), $C_{m1}$ passes only $2N-1$ Grassmann factors. Then by
moving $B_{m1}^{*}$ from left to right, and simultaneously moving $C_{m1}$
from right to left, we can assign to the Grassmann variable in $C_{m1}$ an
additional minus sign comparing with the Grassmann variable in $B_{m1}^{*}$,
and hence obtain the relation of equation (\ref{am2}).

Accordingly, we interchange $B^{*}$ and $C^{*}A^{*}$ in equation (\ref{bw})
to obtain the arrangement of $C^{*}A^{*}B^{*}$ according to the identity
\cite{plechko1}
\begin{equation}
\fl B^{+}\left( CA\right) ^{+}=\frac 12\left[ \left( CA\right)
^{+}B^{+}+\left( CA\right) ^{+}B^{-}+\left( CA\right) ^{-}B^{+}-\left(
CA\right) ^{-}B^{-}\right] ,  \label{exchange}
\end{equation}
with superscripts $+$ and $-$ being the sign factors in boundary Grassmann
factors $A_{1n}^{*}$, $B_{m1}^{*}$ and $C_{1n}^{*}$, and simultaneously move
$C_{m1}$ from right of $\prod\limits_{n=2}^N\stackrel{n}{\overleftarrow{%
C_{mn}A_{mn}}}$ to left of $A_{m1}$ in equations (\ref{bw}) and (\ref{gw}).
Here we note that the superscripts $+$ and $-$ respectively correspond to
periodic and antiperiodic boundary condition imposed on the spin variables
and in turn on the Grassmann variables. Hence, the reduced partition
function becomes
\begin{equation}
\tilde Q=\frac 12\left( \left. \tilde Q_\gamma \right| _{\Gamma _1}+\left.
\tilde Q_\gamma \right| _{\Gamma _2}+\left. \tilde Q_\gamma \right| _{\Gamma
_3}-\left. \tilde Q_\gamma \right| _{\Gamma _4}\right) ,  \label{pppart}
\end{equation}
with
\begin{equation}
\fl \tilde Q_\gamma =\mathop{\rm Sp}\limits_{\left( a,b,c\right) }\left\{ %
\mathop{\rm Sp}\limits_{\left( \sigma \right) }\left[ \left(
\prod_{m=1}^{M-1}\stackrel{m}{\overrightarrow{\Theta _m\Theta _{m+1}^{*}}}%
\right) \cdot \Psi _\gamma \cdot \left(
\prod_{m=1}^M\prod_{n=1}^{N-1}B_{mn}B_{mn+1}^{*}\right) \right] \right\} ,
\end{equation}
and
\begin{equation}
\Psi _\gamma =\mathop{\rm Sp}\limits_{\left( a,b,c\right) }\left\{ %
\mathop{\rm Sp}\limits_{\left( \sigma \right) }\left[ \Theta _1^{*}\left(
\prod_{m=1}^M\stackrel{m}{\overrightarrow{B_{m1}^{*}}}\right) \Theta
_M\left( \prod_{m=1}^M\stackrel{m}{\overleftarrow{B_{mN}}}\right) \right]
\right\} .
\end{equation}
where we have defined
\begin{equation}
\Theta _m=\prod_{n=1}^N\stackrel{n}{\overrightarrow{C_{mn}A_{mn}}}\quad
\mathrm{and\quad }\Theta _m^{*}=\prod_{n=1}^N\stackrel{n}{\overleftarrow{%
C_{mn}^{*}A_{mn}^{*}}},
\end{equation}
and the boundary conditions $\Gamma _1$, $\Gamma _2$, $\Gamma _3$, $\Gamma
_4 $ are defined as
\begin{eqnarray}
\Gamma _1 &=&\left( a_{0n}^{*}=-a_{Mn}^{*},\ b_{m0}^{*}=-b_{mN}^{*},\
c_{0n}^{*}=-c_{Mn}^{*}\right) ,  \label{bc1} \\
\Gamma _2 &=&\left( a_{0n}^{*}=-a_{Mn}^{*},\ b_{m0}^{*}=+b_{mN}^{*},\
c_{0n}^{*}=-c_{Mn}^{*}\right) ,  \label{bc2} \\
\Gamma _3 &=&\left( a_{0n}^{*}=+a_{M,n}^{*},\ b_{m0}^{*}=-b_{mN}^{*},\
c_{0n}^{*}=+c_{M,n}^{*}\right) ,  \label{bc3} \\
\Gamma _4 &=&\left( a_{0n}^{*}=+a_{M,n}^{*},\ b_{m0}^{*}=+b_{mN}^{*},\
c_{0n}^{*}=+c_{M,n}^{*}\right) .  \label{bc4}
\end{eqnarray}
In this way, the configurations of the reduced partition function can be
further rearrangement and expressed as
\begin{eqnarray}
\fl \tilde Q_\gamma &=&\mathop{\rm Sp}\limits_{\left( a,b,c\right) }%
\mathop{\rm Sp}\limits_{\left( \sigma \right) }\left\{ \left(
\prod_{m=1}^{M-1}\stackrel{m}{\overrightarrow{\Theta _m\Theta _{m+1}^{*}}}%
\right) \Theta _1^{*}\left( \prod_{m=1}^M\stackrel{m}{\overrightarrow{%
B_{m1}^{*}}}\right) \Theta _M\left( \prod_{m=1}^M\stackrel{n}{\overleftarrow{%
B_{mN}}}\right) \prod_{m=1}^M\prod_{n=1}^{N-1}B_{mn}B_{mn+1}^{*}\right\}
\nonumber \\
\fl &=&\mathop{\rm Sp}\limits_{\left( a,b,c\right) }\mathop{\rm Sp}%
\limits_{\left( \sigma \right) }\left\{ \left( \prod_{m=1}^M\stackrel{m}{%
\overrightarrow{\Theta _m^{*}B_{m1}^{*}\Theta _m}}\right) \left(
\prod_{m=1}^M\stackrel{n}{\overleftarrow{B_{mN}}}\right) \left(
\prod_{m=1}^M\prod_{n=1}^{N-1}B_{mn}B_{mn+1}^{*}\right) \right\} .
\end{eqnarray}

To have a complete mirror-ordered form, we have to rearrange the terms in
the last two brackets. To achieve this, first we note that
\begin{eqnarray}
\fl \tilde Q_\gamma &=&\mathop{\rm Sp}\limits_{\left( a,b,c\right) }%
\mathop{\rm Sp}\limits_{\left( \sigma \right) }\left\{ \prod_{m=1}^M%
\stackrel{m}{\overrightarrow{\Theta _m^{*}B_{m1}^{*}\left( \prod_{n=1}^{N-1}%
\stackrel{n}{\overrightarrow{C_{mn}A_{mn}}}\right) C_{mN}A_{mN}}}\left(
\prod_{m=1}^M\stackrel{m}{\overleftarrow{B_{mM}}}\right)
\prod_{m=1}^M\prod_{n=1}^{N-1}B_{mn}B_{mn+1}^{*}\right\}  \nonumber \\
\fl &=&\mathop{\rm Sp}\limits_{\left( a,b,c\right) }\mathop{\rm Sp}%
\limits_{\left( \sigma \right) }\left\{ \prod_{m=1}^M\stackrel{m}{%
\overrightarrow{\Theta _m^{*}\left( \prod_{n=1}^{N-1}\stackrel{n}{%
\overrightarrow{B_{mn}^{*}C_{mn}A_{mn}B_{mn}}}\right) B_{mN}^{*}C_{mN}A_{mN}}%
}\left( \prod_{m=1}^M\stackrel{m}{\overleftarrow{B_{mN}}}\right) \right\} .
\end{eqnarray}
The boundary term of with $m=M$, denoted by $T$, can be formulated
\begin{eqnarray}
T &=&\Theta _M^{*}\left( \prod_{n=1}^{N-1}\stackrel{n}{\overrightarrow{%
B_{Mn}^{*}C_{Mn}A_{Mn}B_{Mn}}}\right) B_{MN}^{*}C_{MN}A_{MN}B_{MN}  \nonumber
\\
&=&\left( \prod_{n=1}^N\stackrel{n}{\overleftarrow{C_{Mn}^{*}A_{Mn}^{*}}}%
\right) \left( \prod_{n=1}^{L_y}\stackrel{n}{\overrightarrow{%
B_{Mn}^{*}C_{Mn}A_{Mn}B_{Mn}}}\right)  \nonumber \\
\ &=&\prod_{n=1}^NC_{Mn}^{*}A_{Mn}^{*}B_{Mn}^{*}C_{Mn}A_{Mn}B_{Mn},
\end{eqnarray}
due to the fact that $\mathop{\rm Sp}\limits_{\left( \sigma _{mn}\right)
}\left[ C_{Mn}^{*}A_{Mn}^{*}B_{Mn}^{*}C_{Mn}A_{Mn}B_{Mn}\right] $ for a
given $n$\ is a commutable object. By continuing such construction from $m=M$
down to $m=1$, we can obtain the expression
\begin{equation}
\tilde Q_\gamma =\mathop{\rm Sp}\limits_{\left( a,b,c\right) }\left\{
\prod_{m=1}^M\prod_{n=1}^N\mathop{\rm Sp}\limits_{\left( \sigma _{mn}\right)
}\left[ C_{mn}^{*}A_{mn}^{*}B_{mn}^{*}C_{mn}A_{mn}B_{mn}\right] \right\} .
\end{equation}

For this partition function, the factors containing the same spin are
grouped together and we can perform the average over spins. As a result, we
have
\begin{equation}
\fl \tilde Q_\gamma =\int
\prod_{m=1}^M%
\prod_{n=1}^Nda_{mn}^{*}da_{mn}db_{mn}^{*}db_{mn}dc_{mn}^{*}dc_{mn}\exp
\left( \sum_{m=1}^M\sum_{n=1}^NF_{mn}\right) ,
\end{equation}
with
\begin{eqnarray}
F_{mn} &=&a_{mn}a_{mn}^{*}+b_{mn}b_{mn}^{*}+c_{mn}c_{mn}^{*}  \nonumber \\
&&+r_1r_3c_{m-1n}^{*}a_{m-1n}^{*}  \nonumber \\
&&+\left( r_3c_{m-1n}^{*}+r_1a_{m-1n}^{*}\right) r_2b_{mn-1}^{*}  \nonumber
\\
&&+\left( r_3c_{m-1n}^{*}+r_1a_{m-1n}^{*}+r_2b_{mn-1}^{*}\right) c_{mn-1}
\nonumber \\
&&+\left( r_3c_{m-1n}^{*}+r_1a_{m-1n}^{*}+r_2b_{mn-1}^{*}+c_{mn-1}\right)
a_{mn}  \nonumber \\
&&+\left(
r_3c_{m-1n}^{*}+r_1a_{m-1n}^{*}+r_2b_{mn-1}^{*}+c_{mn-1}+a_{mn}\right)
b_{mn}.
\end{eqnarray}
Since there is no mix on $a_{mn}$ and $b_{mn}$, the integral in the above
expression can be simplified by integrating out the $a_{mn}$ and $b_{mn}$
fields by means of the identity
\begin{equation}
\int dbda\exp \left( \lambda ab+aL+L^{\prime }b\right) \ =\lambda \exp
\left( \lambda ^{-1}LL^{\prime }\right) ,
\end{equation}
where $a$, $b$ are Grassmann variables, $L$, $L^{\prime }$ are linear
fermionic forms independent of $a$, $b$ and $\lambda $ is a parameter. The
result then becomes
\begin{equation}
\tilde Q_\gamma =\int
\prod_{m=1}^M\prod_{n=1}^Ndg_{mn}^{*}dg_{mn}dc_{mn}^{*}dc_{mn}\exp \left(
\sum_{m=1}^M\sum_{n=1}^NG_{mn}\right) ,
\end{equation}
with
\begin{eqnarray}
\fl G_{mn} &=&c_{mn}c_{mn}^{*}+g_{mn}g_{mn}^{*}  \nonumber \\
\fl &&+r_1r_3c_{m-1n}^{*}g_{m-1n}  \nonumber \\
\fl &&-\left( r_3c_{m-1n}^{*}+r_1g_{m-1n}\right) r_2g_{mn-1}^{*}  \nonumber
\\
\fl &&+\left( r_3c_{m-1n}^{*}+r_1g_{m-1n}-r_2g_{mn-1}^{*}\right) c_{mn-1}
\nonumber \\
\fl &&-\left( r_3c_{m-1n}^{*}+r_1g_{m-1n}-r_2g_{mn-1}^{*}+c_{mn-1}\right)
\left( g_{mn}+g_{mn}^{*}\right) ,  \label{fermionic}
\end{eqnarray}
where we have changed the notations for the fields by $\left(
a_{mn}^{*},b_{mn}^{*}\right) \rightarrow \left( g_{mn},-g_{mn}^{*}\right) $.
This is the pure fermionic representation of the reduced partition function.

\section{Exact Solution}

Next, to carry out the integration, we have to use the technique of Fourier
transform to treat the Grassmann variables which mix together with the
variables at different sites. The Fourier transformation is defined as
\begin{equation}
X_{mn}=\frac 1{\sqrt{MN}}\sum_{p=0}^{M-1}\sum_{q=0}^{N-1}X_{pq}e^{-i\frac{%
2\pi }Mmp}e^{-i\frac{2\pi }Nnq},  \label{fourier1}
\end{equation}
and
\begin{equation}
X_{mn}^{*}=\frac 1{\sqrt{MN}}\sum_{p=0}^{M-1}\sum_{q=0}^{N-1}X_{pq}^{*}e^{i%
\frac{2\pi }Mmp}e^{i\frac{2\pi }Nnq},  \label{fourier2}
\end{equation}
where the variables $X_{mn}$ and $X_{mn}^{*}$ denotes one of the variables $%
\left\{ c_{mn},g_{mn}\right\} $ and $\left\{ c_{mn}^{*},g_{mn}^{*}\right\} $
respectively.

After performing the Fourier transformation, the partition function becomes
\begin{equation}
\tilde Q_\gamma =\prod_{p=0}^{M-1}\prod_{q=0}^{N-1}\int dV_{pq}\exp \left(
H_{pq}\right) ,  \label{int1}
\end{equation}
with the measure $dV_{pq}$ defined as
\begin{equation}
dV_{pq}=dg_{pq}^{*}dg_{pq}dc_{pq}^{*}dc_{pq},
\end{equation}
and the function $H_{pq}$ is given by
\begin{eqnarray}
\fl H_{pq} &=&\left( 1-r_3e^{-i\frac{2\pi }Mp}e^{i\frac{2\pi }Nq}\right)
c_{pq}c_{pq}^{*}  \nonumber \\
\fl &&+\left( r_2-e^{i\frac{2\pi }Nq}\right) c_{pq}g_{pq}^{*}+r_3\left(
r_1-e^{-i\frac{2\pi }Mp}\right) c_{pq}^{*}g_{pq}  \nonumber \\
\fl &&-e^{i\frac{2\pi }Nq}\left( 1+r_1e^{-i\frac{2\pi }Mp}\right)
c_{pq}g_{M-pN-q}-r_3e^{-i\frac{2\pi }Mp}\left( 1+r_2e^{i\frac{2\pi }%
Nq}\right) c_{pq}^{*}g_{M-pN-q}^{*}  \nonumber \\
\fl &&+\left( 1-r_1e^{i\frac{2\pi }Mp}-r_2e^{-i\frac{2\pi }Nq}-r_1r_2e^{i%
\frac{2\pi }Mp}e^{-i\frac{2\pi }Nq}\right) g_{pq}g_{pq}^{*}  \nonumber \\
\fl &&-r_1e^{i\frac{2\pi }Mp}g_{pq}g_{M-pN-q}+r_2e^{-i\frac{2\pi }{L_y}%
q}g_{pq}^{*}g_{M-pN-q}^{*},
\end{eqnarray}
Because $H_{pq}$ contains not only the variables, $X_{pq}$ and $X_{pq}^{*}$%
,\ but also the variables, $X_{M-pN-q}$ and $X_{M-pN-q}^{*}$, instead of
calculating $\tilde Q_\gamma $ it is easier to calculate $\tilde Q_\gamma ^2$
given by
\begin{equation}
\tilde Q_\gamma ^2=\prod_{p=0}^{M-1}\prod_{q=0}^{N-1}\int
dV_{pq}dV_{M-pN-q}\exp \left( H_{pq}+H_{M-pN-q}^{*}\right) .  \label{int2}
\end{equation}
Here $H_{M-pN-q}^{*}$ can be obtained from $H_{pq}$ by replacing $p$ by $M-p$
and $q$ by $N-q$ for the Grassmann variables and replacing the coefficient
in front of Grassmann variables by its complex conjugate. Completing the
integration yields
\begin{equation}
\fl Q_\gamma =\prod_{p=0}^{M-1}\prod_{q=0}^{N-1}\left[ A_0-A_1\cos \frac{%
2\pi p}M-A_2\cos \frac{2\pi q}N-A_3\cos \left( \frac{2\pi p}M-\frac{2\pi q}%
N\right) \right] ^{1/2},
\end{equation}
with
\begin{eqnarray}
A_0 &=&\alpha _0^2+\alpha _1^2+\alpha _2^2+\alpha _3^2, \\
A_1 &=&2\left( \alpha _0\alpha _1-\alpha _2\alpha _3\right) , \\
A_2 &=&2\left( \alpha _0\alpha _2-\alpha _1\alpha _3\right) , \\
A_3 &=&2\left( \alpha _0\alpha _3-\alpha _1\alpha _2\right) ,
\end{eqnarray}
where $\alpha _0$, $\alpha _1$, $\alpha _2$, and $\alpha _3$ are given by
equations (\ref{tricoef}) and (\ref{honcoef}) for pt and hc lattices,
respectively.

\subsection{Periodic-periodic boundary condition}

According to equation (\ref{pppart}), the reduced partition function for
ferromagnetic lattices with pp boundary condition is
\begin{equation}
Q^{pp}=\frac 12\left[ \Omega _{\frac 12,\frac 12}+\Omega _{\frac
12,0}+\Omega _{0,\frac 12}-\mathrm{sgn}\left( \frac{\theta -\theta _c}{%
\theta _c}\right) \Omega _{00}\right] ,
\end{equation}
where the superscript $p$ refers to periodic boundary condition and
\begin{equation}
\fl \Omega _{\mu \nu }=\prod_{p=0}^{M-1}\prod_{q=0}^{N-1}\left[ A_0-A_1\cos %
\case{2\pi \left( p+\mu \right) }M-A_2\cos \case{2\pi \left( q+\nu \right) }%
N-A_3\cos \left( \case{2\pi \left( p+\mu \right) }M-\case{2\pi \left( q+\nu
\right) }N\right) \right] ^{1/2}.  \label{exact}
\end{equation}
The sign factor in front of the last term is a result of the standard
consideration of the Grassmann integral over the zero-mode variable $p=q=0$
for ferromagnetic couplings \cite{plechko1,thesis}. When the integral of
equation (\ref{int2}) is carried out, it is always positive, but this is not
the case for equation (\ref{int1}). There are unpaired terms from zero-mode
in equation (\ref{int1}) under various boundary conditions and they
contribute a sign factor to $Q_4$ for $0\leq t_i\leq 1$. The partition
function for pp boundary condition then becomes
\begin{equation}
\fl Z^{pp}=\frac 122^{N_s}\left[ \prod_{i=1}^{n_b}\cosh \left( \beta
J_i\right) \right] ^{N_s}\left[ \Omega _{\frac 12,\frac 12}+\Omega _{\frac
12,0}+\Omega _{0,\frac 12}-\mathrm{sgn}\left( \frac{\theta -\theta _c}{%
\theta _c}\right) \Omega _{00}\right] .
\end{equation}

Furthermore, the free energy density per $k_BT$ of the system defined by
\begin{equation}
f^{pp}=-\frac 1{N_s}\ln Z^{pp},
\end{equation}
then takes the form
\begin{eqnarray}
f^{pp} &=&-\frac{\left( N_s-1\right) }{N_s}\ln 2-\sum_{i=1}^{n_b}\ln \left[
\cosh \left( \beta J_i\right) \right]  \nonumber \\
&&-\frac 1{N_s}\ln \left[ \Omega _{\frac 12,\frac 12}+\Omega _{\frac
12,0}+\Omega _{0,\frac 12}-\mathrm{sgn}\left( \frac{\theta -\theta _c}{%
\theta _c}\right) \Omega _{00}\right] .
\end{eqnarray}

\subsection{Periodic-antiperiodic boundary condition}

For pa boundary condition, equation (\ref{exchange}) is replaced by
\begin{equation}
\fl B^{-}\left( CA\right) ^{+}=\frac 12\left[ \left( CA\right)
^{+}B^{+}+\left( CA\right) ^{+}B^{-}-\left( CA\right) ^{-}B^{+}+\left(
CA\right) ^{-}B^{-}\right] ,
\end{equation}
and the partition function has the form
\begin{equation}
\fl Z^{pa}=\frac 122^{N_s}\left[ \prod_{i=1}^{n_b}\cosh \left( \beta
J_i\right) \right] ^{N_s}\left[ \Omega _{\frac 12,\frac 12}+\Omega _{\frac
12,0}-\Omega _{0,\frac 12}+\mathrm{sgn}\left( \frac{\theta -\theta _c}{%
\theta _c}\right) \Omega _{00}\right] ,
\end{equation}
where the superscript $a$ refers to antiperiodic boundary condition. The
corresponding free energy density per $k_BT$ is
\begin{eqnarray}
f^{pa} &=&-\frac{\left( N_s-1\right) }{N_s}\ln 2-\sum_{i=1}^{n_b}\ln \left[
\cosh \left( \beta J_i\right) \right]  \nonumber \\
&&-\frac 1{N_s}\ln \left[ \Omega _{\frac 12,\frac 12}+\Omega _{\frac
12,0}-\Omega _{0,\frac 12}+\mathrm{sgn}\left( \frac{\theta -\theta _c}{%
\theta _c}\right) \Omega _{00}\right] .
\end{eqnarray}

\subsection{Antiperiodic-periodic boundary condition}

Similarly, for ap boundary condition, equation (\ref{exchange}) is replaced
by
\begin{equation}
\fl B^{+}\left( CA\right) ^{-}=\frac 12\left[ \left( CA\right)
^{+}B^{+}-\left( CA\right) ^{+}B^{-}+\left( CA\right) ^{-}B^{+}+\left(
CA\right) ^{-}B^{-}\right] ,
\end{equation}
and
\begin{equation}
\fl Z^{ap}=\frac 122^{N_s}\left[ \prod_{i=1}^{n_b}\cosh \left( \beta
J_i\right) \right] ^{N_s}\left[ \Omega _{\frac 12,\frac 12}-\Omega _{\frac
12,0}+\Omega _{0,\frac 12}+\mathrm{sgn}\left( \frac{\theta -\theta _c}{%
\theta _c}\right) \Omega _{00}\right] .
\end{equation}
The corresponding free energy density per $k_BT$ is
\begin{eqnarray}
f^{ap} &=&-\frac{\left( N_s-1\right) }{N_s}\ln 2-\sum_{i=1}^{n_b}\ln \left[
\cosh \left( \beta J_i\right) \right]  \nonumber \\
&&-\frac 1{N_s}\ln \left[ \Omega _{\frac 12,\frac 12}-\Omega _{\frac
12,0}+\Omega _{0,\frac 12}+\mathrm{sgn}\left( \frac{\theta -\theta _c}{%
\theta _c}\right) \Omega _{00}\right] .
\end{eqnarray}

\subsection{Antiperiodic-antiperiodic boundary condition}

For aa boundary condition, equation (\ref{exchange}) becomes
\begin{equation}
\fl B^{-}\left( CA\right) ^{-}=\frac 12\left[ -\left( CA\right)
^{+}B^{+}+\left( CA\right) ^{+}B^{-}+\left( CA\right) ^{-}B^{+}+\left(
CA\right) ^{-}B^{-}\right] ,
\end{equation}
and the partition function is
\begin{equation}
\fl Z^{aa}=\frac 122^{N_s}\left[ \prod_{i=1}^{n_b}\cosh \left( \beta
J_i\right) \right] ^{N_s}\left[ -\Omega _{\frac 12,\frac 12}+\Omega _{\frac
12,0}+\Omega _{0,\frac 12}+\mathrm{sgn}\left( \frac{\theta -\theta _c}{%
\theta _c}\right) \Omega _{00}\right] .
\end{equation}
The corresponding free energy density per $k_BT$ is
\begin{eqnarray}
f^{aa} &=&-\frac{\left( N_s-1\right) }{N_s}\ln 2-\sum_{i=1}^{n_b}\ln \left[
\cosh \left( \beta J_i\right) \right]  \nonumber \\
&&-\frac 1{N_s}\ln \left[ -\Omega _{\frac 12,\frac 12}+\Omega _{\frac
12,0}+\Omega _{0,\frac 12}+\mathrm{sgn}\left( \frac{\theta -\theta _c}{%
\theta _c}\right) \Omega _{00}\right] .
\end{eqnarray}

Note that by taking $t_3=0$, $n_b=2$ and $N_s=N_b=MN$, we have $A_3=0$,
\begin{equation}
\Omega _{\mu \nu }=\prod_{p=0}^{M-1}\prod_{q=0}^{N-1}\left[ A_0-A_1\cos %
\case{2\pi \left( p+\mu \right) }M-A_2\cos \case{2\pi \left( q+\nu \right) }%
N\right] ^{1/2},
\end{equation}
and all the results we obtained reduce to those of sq lattice.

Accordingly, the critical temperature can be determined in the thermodynamic
limit from the zero of the free energy contributed by the zero mode,
\begin{equation}
A_0-A_1-A_2-A_3=0.
\end{equation}
It follows that for isotropic coupling, we have
\begin{equation}
\theta _c=\left[ \frac 12\ln \left( 1+\sqrt{2}\right) \right]
^{-1}=2.269185...,
\end{equation}
for sq lattice with $\theta =\left. k_BT\right/ J$,
\begin{equation}
\theta _c=\left[ \frac 12\ln \left( \sqrt{3}\right) \right]
^{-1}=3.640956...,
\end{equation}
for pt lattice and
\begin{equation}
\theta _c=\left[ \frac 12\ln \left( 2+\sqrt{3}\right) \right]
^{-1}=1.518651...,
\end{equation}
for hc lattice.

\section{Specific heat}

The specific heat per spin $C/k_B$ for the Ising model on $M\times N$ sq, pt
and hc lattices with isotropic couplings are shown, respectively, in figure
2(a), 3(a), 4(a) for $M/N=1$, and in figure 2(b), 3(b), 4(b) for $M/N=1/2$.
Figure 3(c) and 4(c) show, respectively, results for pt and hc lattices
under pa and aa boundary conditions and for $M/N=1,1/2,1/4$. In general, for
three lattices with the same lattice size, the specific heat under $pp$
boundary condition is always larger than those under other boundary
conditions. Note that for sq lattices with $M/N=1$, $C^{pa}$ and $C^{aa}$
are distinct in figure 2(a), but for pt and hc lattices with $M/N=1$ in
figure 3(a) and 4(a), they coincide and are non-distinguishable due to the
last term in the bracket of equation (\ref{exact}), which is associated with
the structure symmetries of pt and hc lattices. These behaviours can be
violated by taking aspect ratio $\xi =M/N\neq 1$, and the results are shown
in figures 3(b), (c) and 4(b), (c).

We further study the displacements of the maximum of $C^{pp}$ and $C^{pa}$.
The shift behaviours of the maximum in $C_{NN}\left( T\right) $ are shown in
figure 5. The slopes of the curves implies the rates of approach of $C^{pp}$
and $C^{pa}$ to their limiting behaviours. For periodic-periodic boundary
condition, these lattices have linear behaviours in $N\rightarrow \infty $
and can be described by the formula \cite{ffpr},
\begin{equation}
\frac{\left( T_c-T_{\max }\right) }{T_c}\sim \frac aN,\quad \mathrm{as}\
N\rightarrow \infty .  \label{shift}
\end{equation}
For periodic-antiperiodic boundary condition, the corresponding formula is
also provided by finite-size scaling ansatz. However, for numerical
analysis, instead of equation (\ref{shift}), we use
\begin{equation}
\frac{\left( T_c-T_{\max }\right) }{T_c}=\frac aN+\frac{b_1}{N^2}+\frac{b_2}{%
N^3}+....
\end{equation}
As a result, we have $a_s^{pp}=0.360$, $b_{1,s}^{pp}=-0.47$, $a_s^{pa}=0.18$, $b_{1,s}^{pa}=-2.19$,  for the sq lattice, $%
a_t^{pp}=0.363$, $b_{1,t}^{pp}=-0.91$, $a_t^{pa}=0.09$, $b_{1,t}^{pa}=0.60$ for the pt lattice, $a_h^{pp}=0.268$, $b_{1,h}^{pp}=0.24$, $%
a_h^{pa}=0.09$, $b_{1,h}^{pa}=0.87$ for the hc lattice, and the value of $b_2$ is of the order of $1$. The values of $a^{pp}$ is larger than $%
a^{pa}$ for three lattices, and this implies the approach to
limiting behaviour for $pp$ boundary condition is faster than $pa$
boundary conditions. Since the logarithmic divergence of the
specific heat is independent of boundary conditions and can not be
used to distinguish $C^{pp}$ and $C^{pa}$ of large lattice, then
the values of $a$ may be used to distinguish two boundary
conditions.

\section{Discussion}

We have solved the exact partition functions of $M\times N$ pt and hc
lattices with different boundary conditions. These results can provide the
analytical background for further studies on the effects of lattice
structures and boundary conditions on critical properties and critical
finite-size corrections of the Ising model.

Firstly, universal finite-size scaling functions for critical systems have
received much attention in recent years \cite
{hu,ckhuprl96,okabe,okabepre99,tomita,hphsu,watanabe}, and it is well known
that the finite-size scaling functions depend on the boundary conditions
\cite{huck94}. Hu, Lin and Chen, and Okabe and Kikuchi have discussed the
difference in the finite-size scaling functions for lattice models under
periodic boundary and free boundary conditions in connection with the
universal finite-size scaling function for percolation problem \cite{hu} and
Ising model \cite{okabe}, respectively. Other boundary conditions, such as
the Ising model on an $M\times N$ simple-quartic lattice embedded on a
M\"obius strip and Klein bottle also has been studied \cite{fywu}. Kaneda
and Okabe found that there is interesting aspect ratio dependence of the
value of the Binder parameter at criticality for various boundary conditions
\cite{ko}. It is then interesting to have a rigorous test of finite-size
scaling function and critical finite-size corrections for different planar
Ising model under various boundaries.

In addition, by using Monte Carlo method, Hu, Lin and Chen \cite
{hu,ckhuprl96}, Tomita, Okabe and Hu \cite{tomita} have found that the
universal finite-size scaling functions of the scaled quantities for sq, pt
and hc lattices depend on the aspect ratios and have very good universal
finite-size scaling behaviours when the aspect ratios of these lattices have
the proportions $1:\sqrt{3}/2:\sqrt{3}$. This further implies
lattice-structure-dependence of the universal finite-size scaling function
and it would be a rigorous test from analytical aspect. To aforementioned
topics, we have found finite-size scaling behaviours for sq, pt and hc
lattices under period-aperiodic boundary conditions. By selecting a very
small numbers of nonuniversal metric factors, we have further found very
good universal finite-size scaling behaviours for these lattices, and the
results will be presented in other paper.

Finally, the discussion of the specific heat in this paper also inspire
another problem. Quite recently, Izmailian and Hu have found exact amplitude
ratio and finite-size corrections for the $M\times N$ sq lattice Ising mode
on a torus \cite{izmailian}, and new sets of the universal amplitude ratios
of subdominant correction to scaling amplitudes \cite{izmailianletter}. The
results of section 4 suggest that $a^{pp}/a^{pa}$ for sq, pt and hc lattices
are roughly $2$, $4$, and $3$. The question is $^{\backprime\backprime}$is
there exact relations between $a^{pp}$ and $a^{pa}$ for these lattices ?$%
^{\prime\prime}$ It is interesting to study this question and to have a
heuristic argument for this simple relation.

\ack{The authors wish to thank V. N. Plechko for valuable discussions and a
critical reading of the paper. This work was supported in part by the
National Science Council of the Republic of China (Taiwan) under Grant No.
NSC 90-2112-M-001-074.}

\newpage
\Figure{(a)The global structure of the triangular lattice used in this
paper. A basic cell of the lattice site is given by $(m,n)$, and the
coupling constants are $J_1$, $J_2$ and $J_3$. (b)The global structure of
the honeycomb lattice used in this paper. Each basic cell contains an inner
Ising spin $\sigma_0$.}

\Figure{The specific heat per spin for (a) $N\times N$ square Ising lattices
with isotropic couplings under $pp$, $pa$, $ap$ and $aa$ boundary
conditions, and (b) $M\times N$ square Ising lattices with isotropic
couplings and aspect ratio $M/N=1/2$ under $pp$, $pa$, $ap$ and $aa $
boundary conditions. The critical point $\theta _c$ is marked by a vertical
line. }

\Figure{The specific heat per spin for (a) $N\times N$ plane-triangular
Ising lattices with isotropic couplings under $pp$, $pa$, $ap$ and $aa$
boundary conditions, (b) $M\times N$ plane-triangular Ising lattices with
isotropic couplings and aspect ratio $M/N=1/2$ under $pp$, $pa$, $ap$ and $%
aa $ boundary conditions, and (c) $M\times N$ plane-triangular Ising
lattices with isotropic couplings and aspect ratio $M/N=1,1/2,1/4$ under $pa$
and $aa$ boundary conditions. The critical point $\theta _c$ is marked by a
vertical line.}

\Figure{The specific heat per spin for (a) $N\times N$ honeycomb Ising
lattices with isotropic couplings under $pp$, $pa$, $ap$ and $aa$ boundary
conditions, (b) $M\times N$ honeycomb Ising lattices with isotropic
couplings and aspect ratio $M/N=1/2$ under $pp$, $pa$, $ap$ and $aa$
boundary conditions, and (c) $M\times N$ honeycomb Ising lattices with
isotropic couplings and aspect ratio $M/N=1,1/2,1/4$ under $pa$ and $aa$
boundary conditions. The critical point $\theta _c$ is marked by a vertical
line.}

\Figure{(a) Variation of $\left( T_{\max }-T_c\right) $ with finite $N$ for $%
N\times N$ square Ising lattices with isotropic couplings under $pp$ and $pa$
boundary conditions. The broken lines are given by $\left( T_{\max
}-T_c\right) /T_c=a/N$ and indicate the limiting behaviour as $N\rightarrow
\infty $. (b) Variation of $\left( T_{\max }-T_c\right) $ with finite $N$
for $N\times N$ plane-triangular Ising lattices with isotropic couplings
under $pp$ and $pa$ boundary conditions. (c) Variation of $\left( T_{\max
}-T_c\right) $ with finite $N$ for $N\times N$ honeycomb Ising lattices with
isotropic couplings under $pp$ and $pa$ boundary conditions.}

\end{document}